\documentclass[12pt,reqno]{amsart}
\usepackage{graphicx}
\usepackage{amscd}
\usepackage{amsmath}
\usepackage{epsfig}
\usepackage{amsfonts}
\usepackage{amssymb}

\providecommand{\U}[1]{\protect\rule{.1in}{.1in}}
\providecommand{\U}[1]{\protect\rule{.1in}{.1in}}
\allowdisplaybreaks

\theoremstyle{plain}

\numberwithin{equation}{section}

\input{tcilatex}

\begin{document}
\title[Propagator of a Particle with a Spin]{Propagator of a Charged
Particle with a Spin in \\
Uniform Magnetic and Perpendicular Electric Fields}
\author{Ricardo Cordero--Soto}
\address{Department of Mathematics and Statistics, Arizona State University,
Tempe, AZ 85287--1804, U.S.A.}
\email{ricardojavier81@gmail.com}
\author{Raquel M.~Lopez}
\address{Department of Mathematics and Statistics, Arizona State University,
Tempe, AZ 85287--1804, U.S.A.}
\email{rlopez14@asu.edu}
\author{Erwin Suazo}
\address{Department of Mathematics and Statistics, Arizona State University,
Tempe, AZ 85287--1804, U.S.A.}
\email{suazo@mathpost.la.asu.edu}
\author{Sergei K. Suslov}
\address{Department of Mathematics and Statistics, Arizona State University,
Tempe, AZ 85287--1804, U.S.A.}
\email{sks@asu.edu}
\urladdr{http://hahn.la.asu.edu/\symbol{126}suslov/index.html}
\date{\today }
\dedicatory{Dedicated to the memory of Professor Basil Nicolaenko}

\begin{abstract}
We construct an explicit solution of the Cauchy initial value problem for
the time-dependent Schr\"{o}dinger equation for a charged particle with a
spin moving in a uniform magnetic field and a perpendicular electric field
varying with time. The corresponding Green function (propagator) is given in
terms of elementary functions and certain integrals of the fields with a
characteristic function, which should be found as an analytic or numerical
solution of the equation of motion for the classical oscillator with a
time-dependent frequency. We discuss a particular solution of a related
nonlinear Schr\"{o}dinger equation and some special and limiting cases are
outlined.
\end{abstract}

\maketitle

\section{Introduction}

The time-dependent Schr\"{o}dinger equation%
\begin{equation}
i\hslash \frac{\partial \psi }{\partial t}=H\left( t\right) \psi
\label{int1}
\end{equation}%
can be solved using the time evolution operator given formally by%
\begin{equation}
U\left( t,t_{0}\right) =\text{T}\left( \exp \left( -\frac{i}{\hslash }%
\int_{t_{0}}^{t}H\left( t^{\prime }\right) \ dt^{\prime }\right) \right) ,
\label{int2}
\end{equation}%
where T is the time ordering operator which orders operators with larger
times to the left \cite{Bo:Shi}, \cite{Flu}. This unitary operator takes a
state at time $t_{0}$ to a state at time $t,$ so that%
\begin{equation}
\psi \left( x,t\right) =U\left( t,t_{0}\right) \psi \left( x,t_{0}\right) .
\label{int3}
\end{equation}%
The simplicity of these formulas is deceptive, since the time evolution
operator can be found explicitly as certain integral operator only in a few
special cases. An important example of this source is the forced harmonic
oscillator originally considered by Richard Feynman in his path integrals
approach to the nonrelativistic quantum mechanics \cite{FeynmanPhD}, \cite%
{Feynman}, \cite{Feynman49a}, \cite{Feynman49b}, and \cite{Fey:Hib}; see
also \cite{Lop:Sus}. Since then this problem and its special and limiting
cases were discussed by many authors; see Ref.~\cite{Beauregard}, \cite%
{Gottf:T-MY}, \cite{Holstein}, \cite{Maslov:Fedoriuk}, \cite{Merz}, \cite%
{Thomber:Taylor} for the simple harmonic oscillator and Ref.~\cite%
{Arrighini:Durante}, \cite{Brown:Zhang}, \cite{Holstein97}, \cite{Nardone}, 
\cite{Robinett} for the particle in a constant external field and references
therein. It is worth noting that an exact solution of the $n$-dimensional
time-dependent Schr\"{o}dinger equation for certain modified oscillator is
found in \cite{Me:Co:Su}.\smallskip

In this Letter we construct the time evolution operator explicitly in a
general case of the one-dimensional Schr\"{o}dinger equation when the
Hamiltonian is an arbitrary quadratic form of the operator of coordinate and
the operator of linear momentum; see equation (\ref{schr1}) below. In this
approach, all exactly solvable models, that have been cited above, are
classified in terms of elementary solutions of a characterization equation
given by (\ref{schr13}) below. A particular solution of the corresponding
nonlinear Schr\"{o}dinger equation is obtained in a similar fashion. By
separation of variables, we apply this method to another classical problem
--- the motion of a charged particle with a spin in uniform magnetic and
perpendicular electric fields that are changing with time. The corresponding
Green function (or Feynman's propagator) is found as an elementary function
of certain integrals of our characteristic function and the electromagnetic
fields; see equation (\ref{green3D}) below. Special cases of constant and
linear magnetic fields are discussed as examples. Moreover, these explicit
solutions can also be useful when testing numerical methods of solving the
time-dependent Schr\"{o}dinger equation.

\section{Solution of a Cauchy Initial Value Problem}

The fundamental solution of the time-dependent Schr\"{o}dinger equation with
the quadratic Hamiltonian of the form%
\begin{equation}
i\frac{\partial \psi }{\partial t}=-a\left( t\right) \frac{\partial ^{2}\psi 
}{\partial x^{2}}+b\left( t\right) x^{2}\psi -i\left( c\left( t\right) x%
\frac{\partial \psi }{\partial x}+d\left( t\right) \psi \right) -f\left(
t\right) x\psi +ig\left( t\right) \frac{\partial \psi }{\partial x},
\label{schr1}
\end{equation}%
where $a\left( t\right) ,$ $b\left( t\right) ,$ $c\left( t\right) ,$ $%
d\left( t\right) ,$ $f\left( t\right) ,$ and $g\left( t\right) $ are given
real-valued functions of time $t$ only, can be found with the help of a
familiar substitution%
\begin{equation}
\psi =Ae^{iS}=A\left( t\right) e^{iS\left( x,y,t\right) }  \label{schr2}
\end{equation}%
with%
\begin{equation}
A=A\left( t\right) =\frac{1}{\sqrt{2\pi i\mu \left( t\right) }}
\label{schr3}
\end{equation}%
and%
\begin{equation}
S=S\left( x,y,t\right) =\alpha \left( t\right) x^{2}+\beta \left( t\right)
xy+\gamma \left( t\right) y^{2}+\delta \left( t\right) x+\varepsilon \left(
t\right) y+\kappa \left( t\right) ,  \label{schr4}
\end{equation}%
where $\alpha \left( t\right) ,$ $\beta \left( t\right) ,$ $\gamma \left(
t\right) ,$ $\delta \left( t\right) ,$ $\varepsilon \left( t\right) ,$ and $%
\kappa \left( t\right) $ are differentiable real-valued functions of time $t$
only. Indeed,%
\begin{equation}
\frac{\partial S}{\partial t}=-a\left( \frac{\partial S}{\partial x}\right)
^{2}-bx^{2}+fx+\left( g-cx\right) \frac{\partial S}{\partial x}
\label{schr5}
\end{equation}%
by choosing%
\begin{equation}
\frac{\mu ^{\prime }}{2\mu }=a\frac{\partial ^{2}S}{\partial x^{2}}%
+d=2\alpha \left( t\right) a\left( t\right) +d\left( t\right) .
\label{schr6}
\end{equation}%
Equating the coefficients of all admissible powers of $x^{m}y^{n}$ with $%
0\leq m+n\leq 2,$ gives the following system of ordinary differential
equations%
\begin{align}
& \frac{d\alpha }{dt}+b\left( t\right) +2c\left( t\right) \alpha +4a\left(
t\right) \alpha ^{2}=0,  \label{schr7} \\
& \frac{d\beta }{dt}+\left( c\left( t\right) +4a\left( t\right) \alpha
\left( t\right) \right) \beta =0,  \label{schr8} \\
& \frac{d\gamma }{dt}+a\left( t\right) \beta ^{2}\left( t\right) =0,
\label{schr9} \\
& \frac{d\delta }{dt}+\left( c\left( t\right) +4a\left( t\right) \alpha
\left( t\right) \right) \delta =f\left( t\right) +2\alpha \left( t\right)
g\left( t\right) ,  \label{schr10} \\
& \frac{d\varepsilon }{dt}=\left( g\left( t\right) -2a\left( t\right) \delta
\left( t\right) \right) \beta \left( t\right) ,  \label{schr11} \\
& \frac{d\kappa }{dt}=g\left( t\right) \delta \left( t\right) -a\left(
t\right) \delta ^{2}\left( t\right) ,  \label{schr12}
\end{align}%
where the first equation is the familiar Riccati nonlinear differential
equation; see, for example, \cite{Haah:Stein}, \cite{Rainville}, \cite{Wa}
and references therein. Substitution of (\ref{schr6}) into (\ref{schr7})
results in the second order linear equation%
\begin{equation}
\mu ^{\prime \prime }-\tau \left( t\right) \mu ^{\prime }+4\sigma \left(
t\right) \mu =0  \label{schr13}
\end{equation}%
with%
\begin{equation}
\tau \left( t\right) =\frac{a^{\prime }}{a}-2c+4d,\qquad \sigma \left(
t\right) =ab-cd+d^{2}+\frac{d}{2}\left( \frac{a^{\prime }}{a}-\frac{%
d^{\prime }}{d}\right) ,  \label{schr13a}
\end{equation}%
which must be solved subject to the initial data%
\begin{equation}
\mu \left( 0\right) =0,\qquad \mu ^{\prime }\left( 0\right) =2a\left(
0\right) \neq 0  \label{schr13b}
\end{equation}%
in order to satisfy the initial condition for the corresponding Green
function; see the asymptotic formula (\ref{schr21}) below. We shall refer to
equation (\ref{schr13}) as the \textit{characteristic equation} and its
solution $\mu \left( t\right) ,$ subject to (\ref{schr13b}), as the \textit{%
characteristic function.} As the special case (\ref{schr13}) contains the
generalized equation of hypergeometric type, whose solutions are studied in
detail in \cite{Ni:Uv}; see also \cite{An:As:Ro}, \cite{Ni:Su:Uv}, \cite%
{Sus:Trey}, and \cite{Wa}.\smallskip

Thus, the Green function (fundamental solution or propagator) is explicitly
given in terms of the characteristic function%
\begin{equation}
\psi =G\left( x,y,t\right) =\frac{1}{\sqrt{2\pi i\mu \left( t\right) }}\
e^{i\left( \alpha \left( t\right) x^{2}+\beta \left( t\right) xy+\gamma
\left( t\right) y^{2}+\delta \left( t\right) x+\varepsilon \left( t\right)
y+\kappa \left( t\right) \right) }.  \label{schr14}
\end{equation}%
Here%
\begin{equation}
\alpha \left( t\right) =\frac{1}{4a\left( t\right) }\frac{\mu ^{\prime
}\left( t\right) }{\mu \left( t\right) }-\frac{d\left( t\right) }{2a\left(
t\right) },  \label{schr15}
\end{equation}%
\begin{equation}
\beta \left( t\right) =-\frac{1}{\mu \left( t\right) }\ \exp \left(
-\int_{0}^{t}\left( c\left( \tau \right) -2d\left( \tau \right) \right) \
d\tau \right) ,  \label{scr16}
\end{equation}%
\begin{eqnarray}
\gamma \left( t\right)  &=&\frac{a\left( t\right) }{\mu \left( t\right) \mu
^{\prime }\left( t\right) }\ \exp \left( -2\int_{0}^{t}\left( c\left( \tau
\right) -2d\left( \tau \right) \right) \ d\tau \right)   \label{schr17} \\
&&\quad -4\int_{0}^{t}\frac{a\left( \tau \right) \sigma \left( \tau \right) 
}{\left( \mu ^{\prime }\left( \tau \right) \right) ^{2}}\left( \exp \left(
-2\int_{0}^{\tau }\left( c\left( \lambda \right) -2d\left( \lambda \right)
\right) \ d\lambda \right) \right) \ d\tau ,  \notag
\end{eqnarray}%
\begin{eqnarray}
\delta \left( t\right)  &=&\frac{1}{\mu \left( t\right) }\ \exp \left(
-\int_{0}^{t}\left( c\left( \tau \right) -2d\left( \tau \right) \right) \
d\tau \right) \ \int_{0}^{t}\exp \left( \int_{0}^{\tau }\left( c\left(
\lambda \right) -2d\left( \lambda \right) \right) \ d\lambda \right) 
\label{schr18} \\
&&\quad \times \left( \left( f\left( \tau \right) -\frac{d\left( \tau
\right) }{a\left( \tau \right) }g\left( \tau \right) \right) \mu \left( \tau
\right) +\frac{g\left( \tau \right) }{2a\left( \tau \right) }\mu ^{\prime
}\left( \tau \right) \right) \ d\tau ,  \notag
\end{eqnarray}%
\begin{eqnarray}
\varepsilon \left( t\right)  &=&-\frac{2a\left( t\right) }{\mu ^{\prime
}\left( t\right) }\delta \left( t\right) \ \exp \left( -\int_{0}^{t}\left(
c\left( \tau \right) -2d\left( \tau \right) \right) \ d\tau \right) 
\label{schr18a} \\
&&\quad +8\int_{0}^{t}\frac{a\left( \tau \right) \sigma \left( \tau \right) 
}{\left( \mu ^{\prime }\left( \tau \right) \right) ^{2}}\exp \left(
-\int_{0}^{\tau }\left( c\left( \lambda \right) -2d\left( \lambda \right)
\right) \ d\lambda \right) \left( \mu \left( \tau \right) \delta \left( \tau
\right) \right) \ d\tau   \notag \\
&&\quad +2\int_{0}^{t}\frac{a\left( \tau \right) }{\mu ^{\prime }\left( \tau
\right) }\exp \left( -\int_{0}^{\tau }\left( c\left( \lambda \right)
-2d\left( \lambda \right) \right) \ d\lambda \right) \left( f\left( \tau
\right) -\frac{d\left( \tau \right) }{a\left( \tau \right) }g\left( \tau
\right) \right) \ d\tau ,  \notag
\end{eqnarray}%
\begin{eqnarray}
\kappa \left( t\right)  &=&\frac{a\left( t\right) \mu \left( t\right) }{\mu
^{\prime }\left( t\right) }\delta ^{2}\left( t\right) -4\int_{0}^{t}\frac{%
a\left( \tau \right) \sigma \left( \tau \right) }{\left( \mu ^{\prime
}\left( \tau \right) \right) ^{2}}\left( \mu \left( \tau \right) \delta
\left( \tau \right) \right) ^{2}\ d\tau   \label{schr19} \\
&&\quad -2\int_{0}^{t}\frac{a\left( \tau \right) }{\mu ^{\prime }\left( \tau
\right) }\left( \mu \left( \tau \right) \delta \left( \tau \right) \right)
\left( f\left( \tau \right) -\frac{d\left( \tau \right) }{a\left( \tau
\right) }g\left( \tau \right) \right) \ d\tau   \notag
\end{eqnarray}%
with%
\begin{equation}
\delta \left( 0\right) =\frac{g\left( 0\right) }{2a\left( 0\right) },\qquad
\varepsilon \left( 0\right) =-\delta \left( 0\right) ,\qquad \kappa \left(
0\right) =0.  \label{schr20}
\end{equation}%
We have used integration by parts to resolve the singularities of the
initial data. Then the corresponding asymptotic formula is%
\begin{equation}
G\left( x,y,t\right) =\frac{e^{iS\left( x,y,t\right) }}{\sqrt{2\pi i\mu
\left( t\right) }}\rightarrow \frac{1}{\sqrt{4\pi ia\left( 0\right) t}}\exp
\left( i\frac{\left( x-y\right) ^{2}}{4a\left( 0\right) t}\right) \exp
\left( i\frac{g\left( 0\right) }{2a\left( 0\right) }\left( x-y\right)
\right)   \label{schr21}
\end{equation}%
as $t\rightarrow 0^{+}.$ Notice that the first term on the right hand side
is a familiar free particle propagator (cf.~(\ref{sp1}) below).\smallskip 

By the superposition principle, we obtain an explicit solution of the Cauchy
initial value problem%
\begin{equation}
i\frac{\partial \psi }{\partial t}=H\left( t\right) \psi ,\qquad \left. \psi
\left( x,t\right) \right\vert _{t=0}=\psi _{0}\left( x\right)  \label{schr22}
\end{equation}%
on the infinite interval $-\infty <x<\infty $ with the general quadratic
Hamiltonian as in (\ref{schr1}) in the form%
\begin{equation}
\psi \left( x,t\right) =\int_{-\infty }^{\infty }G\left( x,y,t\right) \ \psi
_{0}\left( y\right) \ dy.  \label{schr23}
\end{equation}%
This yields the time evolution operator (\ref{int2}) explicitly as an
integral operator. Properties of similar oscillatory integrals are discussed
in \cite{SteinHarm}.

\section{Some Special Cases}

Now let us consider several elementary solutions of the characteristic
equation (\ref{schr13}); more complicated cases may include special
functions, like Bessel, hypergeometric or elliptic functions \cite{An:As:Ro}%
, \cite{Ni:Uv}, \cite{Rain}, and \cite{Wa}. Among important elementary cases
of our general expressions for the Green function (\ref{schr14})--(\ref%
{schr19}) are the following:\smallskip

When $a=\hslash /2m,$ $b=c=d=f=g=0,$ and $\mu ^{\prime \prime }=0,$ $\mu
=t\hslash /m,$ one gets%
\begin{equation}
G\left( x,y,t\right) =\sqrt{\frac{m}{2\pi i\hslash t}}\ \exp \left( \frac{%
im\left( x-y\right) ^{2}}{2\hslash t}\right) ,  \label{sp1}
\end{equation}%
which is the free particle propagator \cite{Fey:Hib}.\smallskip

For a particle in a constant external field, where $a=\hslash /2m,$ $%
b=c=d=g=0$ and $f=F/\hslash =\ $constant, $\mu =t\hslash /m,$ the propagator
is given by 
\begin{equation}
G\left( x,y,t\right) =\sqrt{\frac{m}{2\pi i\hslash t}}\ \exp \left( \frac{%
im\left( x-y\right) ^{2}}{2\hslash t}\right) \exp \left( \frac{iF\left(
x+y\right) }{2\hslash }t-\frac{iF^{2}}{24\hslash m}t^{3}\right) .
\label{sp2}
\end{equation}%
This case was studied in detail in \cite{Arrighini:Durante}, \cite%
{Brown:Zhang}, \cite{Fey:Hib}, \cite{Holstein97}, \cite{Nardone} and \cite%
{Robinett}. Here we have corrected a typo in \cite{Fey:Hib}; see \cite{Styer}
for a complete list of known errata in the Feynman and Hibbs book.\smallskip

The simple harmonic oscillator with $a=\hslash /2m,$ $b=m\omega
^{2}/2\hslash ,$ $c=d=f=g=0$ and $\mu ^{\prime \prime }+\omega ^{2}\mu =0,$ $%
\mu =\left( \hslash /m\omega \right) \sin \omega t$ has the familiar
propagator of the form 
\begin{equation}
G\left( x,y,t\right) =\sqrt{\frac{m\omega }{2\pi i\hslash \sin \omega t}}\
\exp \left( \dfrac{im\omega }{2\hslash \sin \omega t}\left( \left(
x^{2}+y^{2}\right) \cos \omega t-2xy\right) \right) ,  \label{sp3}
\end{equation}%
which is studied in detail at \cite{Beauregard}, \cite{Gottf:T-MY}, \cite%
{Holstein}, \cite{Maslov:Fedoriuk}, \cite{Merz}, \cite{Thomber:Taylor}. For
an extension to the case of the forced harmonic oscillator including an
extra velocity-dependent term and a time-dependent frequency, see \cite%
{FeynmanPhD}, \cite{Feynman}, \cite{Fey:Hib} and \cite{Lop:Sus}.\smallskip

Furthermore, an exact solution of the $n$-dimensional time-dependent Schr%
\"{o}dinger equation for certain modified oscillator is found in \cite%
{Me:Co:Su}. In the one-dimensional case we get functions%
\begin{equation}
a=\frac{1}{2}\left( 1+\cos 2t\right) ,\qquad b=\frac{1}{2}\left( 1-\cos
2t\right) ,\qquad c=2d=\sin 2t  \label{sp4}
\end{equation}%
and our characteristic equation (\ref{schr13}) takes the form%
\begin{equation}
\mu ^{\prime \prime }+2\tan t\ \mu ^{\prime }-2\mu =0,  \label{sp5}
\end{equation}%
whose elementary solution is%
\begin{equation}
\mu =\cos t\sinh t+\sin t\cosh t,  \label{sp6}
\end{equation}%
which satisfies the initial conditions (\ref{schr13b}). Further, the
corresponding propagator is given by%
\begin{align}
G\left( x,y,t\right) & =\frac{1}{\sqrt{2\pi i\left( \cos t\sinh t+\sin
t\cosh t\right) }}  \label{sp7} \\
& \quad \times \exp \left( \frac{\left( x^{2}-y^{2}\right) \sin t\sinh
t+2xy-\left( x^{2}+y^{2}\right) \cos t\cosh t}{2i\left( \cos t\sinh t+\sin
t\cosh t\right) }\right) ,  \notag
\end{align}%
which was found in \cite{Me:Co:Su} as the special case $n=1$ of a general $n$%
-dimensional expansion of the Green function in hyperspherical harmonics. We
have showed that (\ref{sp7}) is a generalization of the propagator for the
simple harmonic oscillator; see Ref.~\cite{Me:Co:Su} for more details.

\section{A Particular Solution of the Nonlinear Schr\"{o}dinger equation}

The method of solving the equation (\ref{schr1}) can be extended to the
nonlinear Schr\"{o}dinger equation with a general quadratic Hamiltonian of\
the form%
\begin{eqnarray}
&&i\frac{\partial \psi }{\partial t}=-a\left( t\right) \frac{\partial
^{2}\psi }{\partial x^{2}}+b\left( t\right) x^{2}\psi -i\left( c\left(
t\right) x\frac{\partial \psi }{\partial x}+d\left( t\right) \psi \right) 
\label{nlse1} \\
&&\qquad \quad \quad -f\left( t\right) x\psi +ig\left( t\right) \frac{%
\partial \psi }{\partial x}+h\left( t\right) \left\vert \psi \right\vert
^{2s}\psi ,\qquad s\geq 0,  \notag
\end{eqnarray}%
where $a\left( t\right) ,$ $b\left( t\right) ,$ $c\left( t\right) ,$ $%
d\left( t\right) ,$ $f\left( t\right) ,$ $g\left( t\right) ,$ and $h\left(
t\right) $ are certain functions of time $t$ only. A subs\-titution of%
\begin{equation}
\psi =\psi \left( x,t\right) =\frac{e^{i\phi }}{\sqrt{\mu \left( t\right) }}%
\ e^{iS\left( x,y,t\right) },\qquad \phi =\text{constant},  \label{nlse2}
\end{equation}%
where $S=S\left( x,y,t\right) =\alpha \left( t\right) x^{2}+\beta \left(
t\right) xy+\gamma \left( t\right) y^{2}+\delta \left( t\right)
x+\varepsilon \left( t\right) y+\kappa \left( t\right) $ with the same
relation (\ref{schr6}), results in a modified system (\ref{schr7})--(\ref%
{schr12}) below:%
\begin{align}
& \frac{d\alpha }{dt}+b\left( t\right) +2c\left( t\right) \alpha +4a\left(
t\right) \alpha ^{2}=0,  \label{nlse3} \\
& \frac{d\beta }{dt}+\left( c\left( t\right) +4a\left( t\right) \alpha
\left( t\right) \right) \beta =0,  \label{nlse4} \\
& \frac{d\gamma }{dt}+a\left( t\right) \beta ^{2}\left( t\right) =0,
\label{nlse5} \\
& \frac{d\delta }{dt}+\left( c\left( t\right) +4a\left( t\right) \alpha
\left( t\right) \right) \delta =f\left( t\right) +2\alpha \left( t\right)
g\left( t\right) ,  \label{nlse6} \\
& \frac{d\varepsilon }{dt}=\left( g\left( t\right) -2a\left( t\right) \delta
\left( t\right) \right) \beta \left( t\right) ,  \label{nlse7} \\
& \frac{d\kappa }{dt}=g\left( t\right) \delta \left( t\right) -a\left(
t\right) \delta ^{2}\left( t\right) -\frac{h\left( t\right) }{\mu ^{s}\left(
t\right) },  \label{nlse8}
\end{align}%
where only the last equation has an extra term which corresponds to the
nonlinear term in the original Schr\"{o}dinger equation (\ref{nlse1}).
Therefore equations (\ref{schr13})--(\ref{schr14}) solve once again the
Riccati equation (\ref{nlse3}), however; in this case, we would like to use
nonsingular initial conditions%
\begin{equation}
\mu \left( 0\right) \neq 0,\qquad \mu ^{\prime }\left( 0\right) =2\left(
2\alpha \left( 0\right) a\left( 0\right) +d\left( 0\right) \right) \mu
\left( 0\right) .  \label{nlse9}
\end{equation}%
With these conditions, our system (\ref{nlse3})--(\ref{nlse8}) can be solved
again in terms of the characteristic function $\mu \left( t\right) ,$ thus
giving us a particular solution of the nonlinear Schr\"{o}dinger equation (%
\ref{nlse1}) in the form (\ref{nlse2}), corresponding to the initial data%
\begin{equation}
\psi _{0}\left( x\right) =\left. \psi \left( x,t\right) \right\vert _{t=0}=%
\frac{e^{i\phi }}{\sqrt{\mu \left( 0\right) }}\ e^{i\left( \alpha \left(
0\right) x^{2}+\beta \left( 0\right) xy+\gamma \left( 0\right) y^{2}+\delta
\left( 0\right) x+\varepsilon \left( 0\right) y+\kappa \left( 0\right)
\right) }.  \label{nlse10}
\end{equation}%
The details here are left to the reader.\smallskip\ 

In the simplest case,%
\begin{equation}
i\frac{\partial \psi }{\partial t}=-\frac{1}{2}\frac{\partial ^{2}\psi }{%
\partial x^{2}}+h\left\vert \psi \right\vert ^{2s}\psi ,\qquad h=\text{%
constant},\quad s\geq 0,  \label{nlse11}
\end{equation}%
the solution of the characteristic equation $\mu ^{\prime \prime }=0$ is $%
\mu \left( t\right) =\mu _{0}+t\mu _{1},$ $\mu _{0}>0$ and the coefficients
of the quadratic form are given by%
\begin{eqnarray}
&&\alpha \left( t\right) =\frac{\mu _{1}}{2\left( \mu _{0}+t\mu _{1}\right) }%
,\quad \beta \left( t\right) =\frac{\mu _{0}\beta _{0}}{\mu _{0}+t\mu _{1}}%
,\quad \delta \left( t\right) =\frac{\mu _{0}\delta _{0}}{\mu _{0}+t\mu _{1}}%
,  \label{nlse12} \\
&&\gamma \left( t\right) =\gamma _{0}-\frac{\mu _{0}\beta _{0}^{2}t}{2\left(
\mu _{0}+t\mu _{1}\right) },\qquad \qquad \varepsilon \left( t\right)
=\varepsilon _{0}-\frac{\mu _{0}\beta _{0}\delta _{0}t}{\mu _{0}+t\mu _{1}},
\label{nlse13} \\
&&\kappa \left( t\right) =\kappa _{0}-\frac{\mu _{0}\delta _{0}^{2}t}{%
2\left( \mu _{0}+t\mu _{1}\right) }-\frac{h}{\mu _{1}}\xi _{s}\left( t\right)
\label{nlse14}
\end{eqnarray}%
with%
\begin{equation}
\xi _{s}\left( t\right) =\left\{ 
\begin{array}{ll}
\dfrac{1}{\left( 1-s\right) }\left( \left( \mu _{0}+t\mu _{1}\right)
^{1-s}-\mu _{0}^{1-s}\right) , & \text{when }s\neq 1,\bigskip \\ 
\ln \left( 1+\dfrac{t\mu _{1}}{\mu _{0}}\right) , & \text{when }s=1.%
\end{array}%
\right.  \label{nlse15}
\end{equation}%
Now, the limiting case where $\mu _{1}\rightarrow 0$ with $\mu _{0}>0$ is
given by%
\begin{equation}
\psi =\frac{1}{\sqrt{\mu _{0}}}\ e^{i\left( \beta _{0}xy+\left( \gamma
_{0}-\beta _{0}^{2}t/2\right) y^{2}+\delta _{0}x+\left( \varepsilon
_{0}-\beta _{0}\delta _{0}t\right) y+\kappa _{0}-\delta _{0}^{2}t/2-ht/\mu
_{0}^{s}\right) }.  \label{nlse15a}
\end{equation}%
Then $\left\vert \psi \right\vert =1/\sqrt{\mu _{0}}$ is bounded at all
times. Yet, when $\mu _{1}\neq 0,$ one gets%
\begin{equation}
\left\vert \psi \left( x,t\right) \right\vert =\frac{1}{\sqrt{\mu _{0}+t\mu
_{1}}},\qquad t\geq 0,  \label{nlse15b}
\end{equation}%
which is bounded if $\mu _{0}>0$ and $\mu _{1}>0,$ and blows up at a finite
time $t_{0}=-\mu _{0}/\mu _{1}$ if $\mu _{1}<0.$

The same method shows that the Cauchy initial value problem%
\begin{eqnarray}
&&i\frac{\partial \psi }{\partial t}+\frac{1}{2}\frac{\partial ^{2}\psi }{%
\partial x^{2}}=h\left\vert \psi \right\vert ^{2s}\psi ,\qquad h=\text{%
constant},\quad s\geq 0,  \label{nlse15c} \\
&&\left. \psi \right\vert _{t=0}=\delta _{\varepsilon }\left( x-y\right) =%
\frac{1}{\sqrt{2\pi i\varepsilon }}\ \exp \left( \frac{i\left( x-y\right)
^{2}}{2\varepsilon }\right) ,\quad \varepsilon >0  \notag
\end{eqnarray}%
has the classical solution of the form%
\begin{equation}
\psi =G_{\varepsilon }\left( x,y,t\right) =\frac{1}{\sqrt{2\pi i\left(
t+\varepsilon \right) }}\ \exp \left( \frac{i\left( x-y\right) ^{2}}{2\left(
t+\varepsilon \right) }-\frac{ih}{2\pi }\chi _{s}\left( t\right) \right) ,
\label{nlse15d}
\end{equation}%
where%
\begin{equation}
\chi _{s}\left( t\right) =\left\{ 
\begin{array}{ll}
\dfrac{\left( t+\varepsilon \right) ^{1-s}-\varepsilon ^{1-s}}{1-s}, & \text{%
when }s\neq 1,\bigskip \\ 
\ln \left( 1+\dfrac{t}{\varepsilon }\right) , & \text{when }s=1%
\end{array}%
\right.  \label{nlse15e}
\end{equation}%
with $\chi _{s}\left( 0\right) =0.$ Here%
\begin{equation}
\left. \psi \right\vert _{t=0}=G_{\varepsilon }\left( x,y,0\right) =\delta
_{\varepsilon }\left( x-y\right) \rightarrow \delta \left( x-y\right)
\label{nlse15f}
\end{equation}%
as $\varepsilon \rightarrow 0^{+}$ in the distributional sense%
\begin{equation}
\lim_{\varepsilon \rightarrow 0^{+}}\int_{-\infty }^{\infty }G_{\varepsilon
}\left( x,y,0\right) \varphi \left( y\right) \ dy=\varphi \left( x\right) .
\label{nlse15g}
\end{equation}%
Further details are left to the reader.\smallskip

In a similar fashion, the nonlinear Schr\"{o}dinger equation%
\begin{equation}
i\frac{\partial \psi }{\partial t}=-\cos ^{2}t\ \frac{\partial ^{2}\psi }{%
\partial x^{2}}+\sin ^{2}t\ x^{2}\psi -i\sin t\cos t\left( 2x\frac{\partial
\psi }{\partial x}+\psi \right) +2\cos t\sinh t\left\vert \psi \right\vert
^{2s}\psi  \label{nlse16}
\end{equation}%
(cf. \cite{Me:Co:Su}) has a particular solution of the form%
\begin{equation}
\psi =\psi \left( x,t\right) =\frac{1}{\sqrt{\mu \left( t\right) }}\
e^{i\left( \alpha \left( t\right) x^{2}+\beta \left( t\right) xy+\gamma
\left( t\right) y^{2}+\kappa \left( t\right) \right) },  \label{nlse17}
\end{equation}%
where $\mu \left( t\right) =\cos t\cosh t+\sin t\sinh t$ and the
coefficients are given by%
\begin{eqnarray}
&&\alpha \left( t\right) =\frac{\cos t\sinh t-\sin t\cosh t}{2\left( \cos
t\cosh t+\sin t\sinh t\right) },  \label{nlse18} \\
&&\beta \left( t\right) =\frac{1}{\cos t\cosh t+\sin t\sinh t},
\label{nlse19} \\
&&\gamma \left( t\right) =-\frac{\cos t\sinh t+\sin t\cosh t}{2\left( \cos
t\cosh t+\sin t\sinh t\right) },  \label{nlse20} \\
&&\kappa \left( t\right) =\left\{ 
\begin{array}{ll}
-\dfrac{\left( \cos t\cosh t+\sin t\sinh t\right) ^{1-s}-1}{1-s}, & \text{%
when }s\neq 1,\bigskip \\ 
-\ln \left( \cos t\cosh t+\sin t\sinh t\right) , & \text{when }s=1.%
\end{array}%
\right.  \label{nlse21}
\end{eqnarray}%
Thus, initial function is the standing wave%
\begin{equation}
\psi _{0}\left( x\right) =\left. \psi \left( x,t\right) \right\vert
_{t=0}=e^{ixy}.  \label{nlse22}
\end{equation}%
These exact solutions may be of interest in a general treatment of the
nonlinear Schr\"{o}dinger equation (see \cite{Howland}, \cite{Jafaev}, \cite%
{Naibo:Stef}, \cite{Rod:Schlag}, \cite{Schlag}, \cite{Yajima} and references
therein).

\section{Motion in Perpendicular Magnetic and Electric Fields}

A particle with a spin $s$ also has a magnetic momentum $\boldsymbol{\mu }$
with the operator%
\begin{equation}
\widehat{\boldsymbol{\mu }}=\mu \widehat{\boldsymbol{s}}/s,  \label{ll1}
\end{equation}%
where $\widehat{\boldsymbol{s}}$ is the spin operator and $\mu $ is a
constant characterizing the particle. For the motion of a charged particle
in a uniform magnetic field $\boldsymbol{H}$ and an electric field $%
\boldsymbol{E},$ both which are perpendicular to each other, the
three-dimensional time-dependent Schr\"{o}dinger equation%
\begin{equation}
i\hslash \frac{\partial \Psi }{\partial t}=\widehat{H}\Psi  \label{ll2}
\end{equation}%
has a Hamiltonian operator as in \cite{La:Lif}, namely,%
\begin{equation}
\widehat{H}=\frac{1}{2m}\left( \widehat{p}_{x}+\frac{eH}{c}y\right) ^{2}+%
\frac{1}{2m}\widehat{p}_{y}^{2}+\frac{1}{2m}\widehat{p}_{z}^{2}-\frac{\mu }{s%
}\widehat{s}_{z}H-yF,  \label{ll3}
\end{equation}%
where $\widehat{\boldsymbol{p}}=-i\hslash \boldsymbol{\nabla }$ is the
linear momentum operator. The corresponding vector and scalar potentials are
defined up to the gauge transformation. We use the original choice \cite%
{La:Lif} for the vector potential $\boldsymbol{A}=-yH\left( t\right) \ 
\boldsymbol{e}_{x}$ and add a linear scalar potential $\varphi =-\left(
F\left( t\right) /e\right) y$ (see also \cite{Lop:Sus}). Then the uniform
magnetic field $\boldsymbol{H}$ and the corresponding perpendicular electric
field $\boldsymbol{E}$ are given by 
\begin{equation}
\boldsymbol{H}=\func{curl}\boldsymbol{A}=H\left( t\right) \ \boldsymbol{e}%
_{z},\qquad \boldsymbol{E}=-\boldsymbol{\nabla }\varphi -\frac{1}{c}\frac{%
\partial \boldsymbol{A}}{\partial t}=\frac{y}{c}\frac{dH\left( t\right) }{dt}%
\boldsymbol{e}_{x}+\frac{F\left( t\right) }{e}\ \boldsymbol{e}_{y};
\label{ll3a}
\end{equation}%
see \cite{Jack} for more details.\smallskip

Since (\ref{ll3}) does not contain the other components of the spin, the
operator $\widehat{s}_{z}$ commutes with the Hamiltonian $\widehat{H}$ and
the $z$-component of the spin is conserved. Thus the operator $\widehat{s}%
_{z}$ can be replaced by its eigenvalue $s_{z}=\sigma $ in the Hamiltonian (%
\ref{ll3}), so we have%
\begin{equation}
\widehat{H}=\frac{1}{2m}\left( \widehat{p}_{x}+\frac{eH}{c}y\right) ^{2}+%
\frac{1}{2m}\widehat{p}_{y}^{2}+\frac{1}{2m}\widehat{p}_{z}^{2}-\frac{\mu
\sigma }{s}H-yF  \label{ll4}
\end{equation}%
with $\sigma =-s,-s+1,...\ ,s-1,s.$ Then the spin dependence of the wave
function becomes insignificant and the wave function in the Schr\"{o}dinger
equation (\ref{ll2}) can be taken as an ordinary coordinate function $\Psi
=\Psi \left( \boldsymbol{r},t,\sigma \right) .$\smallskip

It should be noted that, the Hamiltonian (\ref{ll4}) does not contain the
coordinates $x$ and $z$ explicitly. Therefore the operators $\widehat{p}_{x}$
and $\widehat{p}_{z}$ also commute with the Hamiltonian and the $x$ and $z$
components of the linear momentum are conserved. The corresponding
eigenvalues $p_{x}$ and $p_{z}$ take on all real values from $-\infty $ to $%
\infty .$ In this Letter we consider the case when the magnetic field $H$
and the electric\ force $F$ are arbitrary functions of time $t$ only. Using
the substitution%
\begin{equation}
\Psi \left( \boldsymbol{r},t\right) =e^{i\left( xp_{x}+zp_{z}-S_{F}\left(
t\right) \right) /\hslash }\ \psi \left( y,t\right)  \label{ll5}
\end{equation}%
with%
\begin{equation}
\frac{dS_{F}}{dt}=\frac{p_{z}^{2}}{2m}-\frac{\mu \sigma }{s}H\left( t\right)
+\frac{cp_{x}}{e}\frac{F\left( t\right) }{H\left( t\right) }  \label{ll6}
\end{equation}%
results in the one-dimensional Schr\"{o}dinger equation of the harmonic
oscillator driven by an external force in the $y$-direction 
\begin{equation}
i\hslash \frac{\partial \psi }{\partial t}=-\frac{\hslash ^{2}}{2m}\frac{%
\partial ^{2}\psi }{\partial y^{2}}+\frac{m\omega _{H}^{2}}{2}\left(
y-y_{0}\right) ^{2}\psi -F\left( t\right) \left( y-y_{0}\right) \psi
\label{ll7}
\end{equation}%
with time-dependent quantities%
\begin{equation}
\omega _{H}\left( t\right) =\frac{\left\vert e\right\vert H\left( t\right) }{%
mc},\qquad y_{0}\left( t\right) =-\frac{cp_{x}}{eH\left( t\right) },\qquad
a_{H}\left( t\right) =\sqrt{\frac{\hslash }{m\omega _{H}\left( t\right) }}.
\label{ll8}
\end{equation}%
When the magnetic field is a constant $H^{\prime }\equiv 0,$ the solution is
well known (see \cite{La:Lif}, \cite{Lop:Sus}). In the absence of external
electric force $F\equiv 0,$ there exist discrete energy values corresponding
to motion in a plane perpendicular to the magnetic field, namely, the 
\textit{Landau levels}. See \cite{La:Lif}, \cite{Lop:Sus} and section~7 for
more details.\smallskip

In general, the following substitution%
\begin{equation}
\psi \left( y,t\right) =\chi \left( \eta ,t\right) ,\qquad \eta =\frac{%
y-y_{0}\left( t\right) }{a_{H}\left( t\right) }  \label{ll9}
\end{equation}%
gives the time-dependent Schr\"{o}dinger equation of the form (\ref{schr1})
as follows,%
\begin{equation}
i\frac{\partial \chi }{\partial t}=\frac{\omega _{H}\left( t\right) }{2}%
\left( -\frac{\partial ^{2}}{\partial \eta ^{2}}+\eta ^{2}\right) \chi
-f\left( t\right) \eta \chi +i\left( g\left( t\right) +h\left( t\right) \eta
\right) \frac{\partial \chi }{\partial \eta },  \label{ll10}
\end{equation}%
where%
\begin{equation}
f\left( t\right) =\frac{a_{H}\left( t\right) F\left( t\right) }{\hslash }%
,\qquad g\left( t\right) =\frac{y_{0}^{\prime }\left( t\right) }{a_{H}\left(
t\right) },\qquad h\left( t\right) =\frac{a_{H}^{\prime }\left( t\right) }{%
a_{H}\left( t\right) },  \label{ll11}
\end{equation}%
which can be solved by the method discussed in the previous sections. In
this case%
\begin{equation*}
\tau \left( t\right) =\frac{\omega _{H}^{\prime }\left( t\right) }{\omega
_{H}\left( t\right) }+2\frac{a_{H}^{\prime }\left( t\right) }{a_{H}\left(
t\right) }=\left( \ln \left( \omega _{H}a_{H}^{2}\right) \right) ^{\prime
}=0,\qquad \sigma \left( t\right) =\frac{1}{4}\omega _{H}^{2}\left( t\right)
\end{equation*}%
and the corresponding characteristic equation (\ref{schr13}) coincides with\
the equation of motion for the classical oscillator with a time-dependent
frequency%
\begin{equation}
\mu _{H}^{\prime \prime }+\omega _{H}^{2}\left( t\right) \mu _{H}=0,\qquad
\mu _{H}\left( 0\right) =0,\quad \mu _{H}^{\prime }\left( 0\right) =\omega
_{H}\left( 0\right) .  \label{ll12}
\end{equation}%
The fundamental solution of (\ref{ll10}) is given by%
\begin{equation}
\chi =G_{H}\left( \eta ,\eta ^{\prime },t\right) =\frac{1}{\sqrt{2\pi i\mu
_{H}\left( t\right) }}\ e^{i\left( \alpha _{H}\left( t\right) \eta
^{2}+\beta _{H}\left( t\right) \eta \eta ^{\prime }+\gamma _{H}\left(
t\right) \left( \eta ^{\prime }\right) ^{2}+\delta _{H}\left( t\right) \eta
+\varepsilon _{H}\left( t\right) \eta ^{\prime }+\kappa _{H}\left( t\right)
\right) },  \label{greeny}
\end{equation}%
where%
\begin{equation}
\alpha _{H}\left( t\right) =\frac{1}{2\omega _{H}\left( t\right) }\frac{d}{dt%
}\left( \ln \mu _{H}\left( t\right) \right) =\frac{m}{2\hslash }%
a_{H}^{2}\left( t\right) \frac{d}{dt}\left( \ln \mu _{H}\left( t\right)
\right) ,  \label{alpha}
\end{equation}%
\begin{equation}
\beta _{H}\left( t\right) =-\frac{1}{\mu _{H}\left( t\right) }\frac{%
a_{H}\left( t\right) }{a_{H}\left( 0\right) }=-\frac{1}{\mu _{H}\left(
t\right) }\sqrt{\frac{\omega _{H}\left( 0\right) }{\omega _{H}\left(
t\right) }}=-\frac{1}{\mu _{H}\left( t\right) }\sqrt{\frac{H\left( 0\right) 
}{H\left( t\right) }},  \label{beta}
\end{equation}%
\begin{equation}
\gamma _{H}\left( t\right) =\frac{\omega _{H}\left( 0\right) }{2}\left( 
\frac{1}{\mu _{H}\left( t\right) \mu _{H}^{\prime }\left( t\right) }%
-\int_{0}^{t}\left( \frac{\omega _{H}\left( \tau \right) }{\mu _{H}^{\prime
}\left( \tau \right) }\right) ^{2}\right) \ d\tau ,  \label{gamma}
\end{equation}%
\begin{equation}
\delta _{H}\left( t\right) =\frac{a_{H}\left( t\right) }{\hslash \mu
_{H}\left( t\right) }\left( \delta _{F}^{\left( 0\right) }\left( t\right)
+p_{x}\delta _{H}^{\left( 1\right) }\left( t\right) \right) ,  \label{delta}
\end{equation}%
\begin{equation}
\delta _{F}^{\left( 0\right) }\left( t\right) =\int_{0}^{t}\mu _{H}\left(
\tau \right) F\left( \tau \right) \ d\tau ,  \label{delta0}
\end{equation}%
\begin{eqnarray}
\delta _{H}^{\left( 1\right) }\left( t\right) &=&\frac{mc}{e}\int_{0}^{t}%
\frac{\mu _{H}^{\prime }\left( \tau \right) H^{\prime }\left( \tau \right) }{%
H^{2}\left( \tau \right) }\ d\tau  \label{delta1} \\
&=&\frac{e}{\left\vert e\right\vert }-\frac{mc\mu _{H}^{\prime }\left(
t\right) }{eH\left( t\right) }-\frac{e}{mc}\int_{0}^{t}\mu _{H}\left( \tau
\right) H\left( \tau \right) \ d\tau ,  \notag
\end{eqnarray}%
\begin{equation}
\varepsilon _{H}\left( t\right) =\frac{1}{ma_{H}\left( 0\right) }\left(
\varepsilon _{F}^{\left( 0\right) }\left( t\right) +p_{x}\varepsilon
_{H}^{\left( 1\right) }\left( t\right) \right) ,  \label{epsilon}
\end{equation}%
\begin{eqnarray}
\varepsilon _{F}^{\left( 0\right) }\left( t\right) &=&\int_{0}^{t}\left( 1-%
\frac{\mu _{H}\left( \tau \right) \mu _{H}^{\prime }\left( \tau \right) }{%
\mu _{H}\left( t\right) \mu _{H}^{\prime }\left( t\right) }\right) \frac{%
F\left( \tau \right) }{\mu _{H}^{\prime }\left( \tau \right) }\ d\tau
\label{epsilon0} \\
&&\quad +\frac{\hslash e^{2}}{m^{2}c^{2}}\int_{0}^{t}\left( \frac{H\left(
\tau \right) }{\mu _{H}^{\prime }\left( \tau \right) }\right) ^{2}\delta
_{F}^{\left( 0\right) }\left( t\right) \ d\tau ,  \notag
\end{eqnarray}%
\begin{equation}
\varepsilon _{H}^{\left( 1\right) }\left( t\right) =-\frac{\delta
_{H}^{\left( 1\right) }\left( t\right) }{\mu _{H}\left( t\right) \mu
_{H}^{\prime }\left( t\right) }+\frac{\hslash e^{2}}{m^{2}c^{2}}%
\int_{0}^{t}\left( \frac{H\left( \tau \right) }{\mu _{H}^{\prime }\left(
\tau \right) }\right) ^{2}\delta _{H}^{\left( 1\right) }\left( \tau \right)
\ d\tau ,  \label{epsilon1}
\end{equation}%
\begin{equation}
\kappa _{H}\left( t\right) =\frac{1}{2\hslash m}\left( \kappa _{F}^{\left(
0\right) }\left( t\right) +p_{x}\kappa _{F}^{\left( 1\right) }\left(
t\right) +p_{x}^{2}\kappa _{H}^{\left( 2\right) }\left( t\right) \right) ,
\label{kappa}
\end{equation}%
\begin{eqnarray}
\kappa _{F}^{\left( 0\right) }\left( t\right) &=&\frac{1}{\mu _{H}\left(
t\right) \mu _{H}^{\prime }\left( t\right) }\left( \delta _{F}^{\left(
0\right) }\left( t\right) \right) ^{2}-\int_{0}^{t}\left( \frac{\omega
_{H}\left( \tau \right) \delta _{F}^{\left( 0\right) }\left( t\right) }{\mu
_{H}^{\prime }\left( \tau \right) }\right) ^{2}\ d\tau  \label{kappa0} \\
&&\quad -2\int_{0}^{t}\frac{F\left( \tau \right) \delta _{H}^{\left(
1\right) }\left( \tau \right) }{\mu _{H}^{\prime }\left( \tau \right) }\
d\tau ,  \notag
\end{eqnarray}%
\begin{eqnarray}
\frac{1}{2}\kappa _{F}^{\left( 1\right) }\left( t\right) &=&\frac{\delta
_{F}^{\left( 0\right) }\left( t\right) \delta _{H}^{\left( 1\right) }\left(
t\right) }{\mu _{H}\left( t\right) \mu _{H}^{\prime }\left( t\right) }%
-\int_{0}^{t}\left( \frac{\omega _{H}\left( \tau \right) }{\mu _{H}^{\prime
}\left( \tau \right) }\right) ^{2}\delta _{F}^{\left( 0\right) }\left(
t\right) \delta _{H}^{\left( 1\right) }\left( t\right) \ d\tau
\label{kappa1} \\
&&\quad -\int_{0}^{t}\frac{F\left( \tau \right) \delta _{H}^{\left( 1\right)
}\left( \tau \right) }{\mu _{H}^{\prime }\left( \tau \right) }\ d\tau , 
\notag
\end{eqnarray}%
\begin{equation}
\kappa _{H}^{\left( 2\right) }\left( t\right) =\frac{\left( \delta
_{H}^{\left( 1\right) }\left( t\right) \right) ^{2}}{\mu _{H}\left( t\right)
\mu _{H}^{\prime }\left( t\right) }-\int_{0}^{t}\left( \frac{\omega
_{H}\left( \tau \right) \delta _{H}^{\left( 1\right) }\left( \tau \right) }{%
\mu _{H}^{\prime }\left( \tau \right) }\right) ^{2}\ d\tau  \label{kappa2}
\end{equation}%
as a special case of (\ref{schr14})--(\ref{schr19}).

\section{The Propagator in Three Dimensions}

By the superposition principle, the fundamental solution $\Psi =G\left( 
\boldsymbol{r},\boldsymbol{r}^{\prime },t\right) $ of the time-dependent Schr%
\"{o}dinger equation of a particle with a spin in a uniform magnetic field%
\begin{equation}
\left( i\hslash \frac{\partial }{\partial t}-\widehat{H}\right) G\left( 
\boldsymbol{r},\boldsymbol{r}^{\prime },t\right) =0  \label{3D1}
\end{equation}%
is given by the double Fourier integral of the form%
\begin{eqnarray}
G\left( \boldsymbol{r},\boldsymbol{r}^{\prime },t\right) &=&\frac{1}{\left(
2\pi \hslash \right) ^{2}a_{H}\left( 0\right) }\diint_{-\infty }^{\infty
}e^{i\left( \left( x-x^{\prime }\right) p_{x}+\left( z-z^{\prime }\right)
p_{z}-S_{F}\left( t,p_{x},p_{z}\right) \right) /\hslash }  \label{3D2} \\
&&\qquad \quad \quad \times \ G_{H}\left( \frac{y-y_{0}\left( t\right) }{%
a_{H}\left( t\right) },\frac{y^{\prime }-y_{0}\left( 0\right) }{a_{H}\left(
0\right) },t\right) \ dp_{x}dp_{z}.  \notag
\end{eqnarray}%
Here we obtain from (\ref{ll6}) that 
\begin{equation}
S_{F}\left( t,p_{x},p_{z}\right) =\frac{p_{z}^{2}}{2m}t+\frac{cp_{x}}{e}%
\int_{0}^{t}\frac{F\left( \tau \right) }{H\left( \tau \right) }\ d\tau -%
\frac{\mu \sigma }{s}\int_{0}^{t}H\left( \tau \right) \ d\tau .  \label{3D3}
\end{equation}%
By (\ref{greeny}) the corresponding Green function is represented as%
\begin{equation}
G_{H}\left( \eta ,\eta ^{\prime },t\right) =\frac{1}{\sqrt{2\pi i\mu
_{H}\left( t\right) }}\ e^{iS_{H}\left( \eta ,\eta ^{\prime },t\right) }
\label{3D4}
\end{equation}%
with%
\begin{equation}
S_{H}\left( \eta ,\eta ^{\prime },t\right) =\alpha _{H}\left( t\right) \eta
^{2}+\beta _{H}\left( t\right) \eta \eta ^{\prime }+\gamma _{H}\left(
t\right) \left( \eta ^{\prime }\right) ^{2}+\delta _{H}\left( t\right) \eta
+\varepsilon _{H}\left( t\right) \eta ^{\prime }+\kappa _{H}\left( t\right)
\label{3D5}
\end{equation}%
and%
\begin{equation*}
\eta =\frac{y-y_{0}\left( t\right) }{a_{H}\left( t\right) },\qquad \eta
^{\prime }=\frac{y^{\prime }-y_{0}\left( 0\right) }{a_{H}\left( 0\right) },
\end{equation*}%
where $y_{0}$ is a linear function of $p_{x}$ [see (\ref{ll8}) for a
definition of functions $y_{0}\left( t\right) $ and $a_{H}\left( t\right) $%
]. Next, we are given%
\begin{equation}
\lim_{t\rightarrow 0^{+}}G\left( \boldsymbol{r},\boldsymbol{r}^{\prime
},t\right) =\delta \left( \boldsymbol{r}-\boldsymbol{r}^{\prime }\right)
=\delta \left( x-x^{\prime }\right) \delta \left( y-y^{\prime }\right)
\delta \left( z-z^{\prime }\right)  \label{dirac3D}
\end{equation}%
as our initial data by the asymptotic relation (\ref{schr21}) and the
integral representation%
\begin{equation}
\delta \left( \alpha \right) =\frac{1}{2\pi }\int_{-\infty }^{\infty
}e^{i\alpha \xi }\ d\xi  \label{dirac}
\end{equation}%
as the Dirac delta function.\smallskip

The quadratic in (\ref{3D5}) is also a quadratic polynomial in $p_{x}$ given
as:%
\begin{equation}
S_{H}\left( \eta ,\eta ^{\prime },t\right) =S_{H}^{\left( 2\right) }\left(
y,y^{\prime },t\right) +p_{x}\ S_{H}^{\left( 1\right) }\left( y,y^{\prime
},t\right) +p_{x}^{2}\ S_{H}^{\left( 0\right) }\left( t\right)  \label{3D6}
\end{equation}%
with coefficients%
\begin{eqnarray}
S_{H}^{\left( 0\right) }\left( t\right) &=&\frac{c^{2}\alpha _{H}\left(
t\right) }{e^{2}a_{H}^{2}\left( t\right) H^{2}\left( t\right) }+\frac{%
c^{2}\beta _{H}\left( t\right) }{e^{2}a_{H}\left( t\right) a_{H}\left(
0\right) H\left( t\right) H\left( 0\right) }+\frac{c^{2}\gamma _{H}\left(
t\right) }{e^{2}a_{H}^{2}\left( 0\right) H^{2}\left( 0\right) }  \label{3D7}
\\
&&\quad +\frac{c\delta _{H}^{\left( 1\right) }\left( t\right) }{\hslash e\mu
_{H}\left( t\right) H\left( t\right) }+\frac{c\varepsilon _{H}^{\left(
1\right) }\left( t\right) }{mea_{H}^{2}\left( 0\right) H\left( 0\right) }+%
\frac{\kappa _{H}^{\left( 2\right) }\left( t\right) }{2\hslash m},  \notag
\end{eqnarray}%
\begin{eqnarray}
S_{H}^{\left( 1\right) }\left( y,y^{\prime },t\right) &=&\left( \frac{%
2c\alpha _{H}\left( t\right) }{ea_{H}^{2}\left( t\right) H\left( t\right) }+%
\frac{c\beta _{H}\left( t\right) }{ea_{H}\left( t\right) a_{H}\left(
0\right) H\left( 0\right) }+\frac{\delta _{H}^{\left( 1\right) }\left(
t\right) }{\hslash \mu _{H}\left( t\right) }\right) y  \label{3D8} \\
&&\quad +\left( \frac{c\beta _{H}\left( t\right) }{ea_{H}\left( t\right)
a_{H}\left( 0\right) H\left( t\right) }+\frac{2c\gamma _{H}\left( t\right) }{%
ea_{H}^{2}\left( 0\right) H\left( 0\right) }+\frac{\varepsilon _{H}^{\left(
1\right) }\left( t\right) }{ma_{H}^{2}\left( 0\right) }\right) y^{\prime } 
\notag \\
&&\quad \quad +\frac{c\delta _{F}^{\left( 0\right) }\left( t\right) }{%
\hslash e\mu _{H}\left( t\right) H\left( t\right) }+\frac{c\varepsilon
_{F}^{\left( 0\right) }\left( t\right) }{mea_{H}^{2}\left( 0\right) H\left(
0\right) }+\frac{\kappa _{F}^{\left( 1\right) }\left( t\right) }{2\hslash m},
\notag
\end{eqnarray}%
and%
\begin{eqnarray}
S_{H}^{\left( 2\right) }\left( y,y^{\prime },t\right) &=&\frac{\alpha
_{H}\left( t\right) }{a_{H}^{2}\left( t\right) }y^{2}+\frac{\beta _{H}\left(
t\right) }{a_{H}\left( t\right) a_{H}\left( 0\right) }yy^{\prime }+\frac{%
\gamma _{H}\left( t\right) }{a_{H}^{2}\left( 0\right) }\left( y^{\prime
}\right) ^{2}  \label{3D9} \\
&&\quad +\frac{\delta _{F}^{\left( 0\right) }\left( t\right) }{\hslash \mu
_{H}\left( t\right) }y+\frac{\varepsilon _{F}^{\left( 0\right) }\left(
t\right) }{ma_{H}^{2}\left( 0\right) }y^{\prime }+\frac{\kappa _{F}^{\left(
0\right) }\left( t\right) }{2\hslash m}.  \notag
\end{eqnarray}%
See equations (\ref{ll8}) and (\ref{alpha})--(\ref{kappa2}) for notation
explanation.\smallskip

Hence, the double Fourier integral (\ref{3D2}) can be evaluated in terms of
elementary functions. So, integration over $p_{z}$ gives the free particle
propagator of motion in the direction of magnetic field in the following
fashion,%
\begin{align}
G_{0}\left( z-z^{\prime },t\right) & =\frac{1}{2\pi \hslash }\int_{-\infty
}^{\infty }\exp \left( \dfrac{i}{\hslash }\left( \left( z-z^{\prime }\right)
p_{z}-\dfrac{p_{z}^{2}}{2m}t\right) \right) \ dp_{z}  \label{3D10} \\
& =\sqrt{\frac{m}{2\pi i\hslash t}}\ \exp \left( \frac{im\left( z-z^{\prime
}\right) ^{2}}{2\hslash t}\right)  \notag
\end{align}%
using the familiar elementary integral%
\begin{equation}
\int_{-\infty }^{\infty }e^{i\left( az^{2}+2bz\right) }\,dz=\sqrt{\frac{\pi i%
}{a}}\,e^{-ib^{2}/a};  \label{gauss}
\end{equation}%
see Refs.~\cite{Bo:Shi} and \cite{Palio:Mead}.\smallskip

Similarly, integration over $p_{x}$ yields%
\begin{eqnarray}
G\left( \boldsymbol{r},\boldsymbol{r}^{\prime },t\right) &=&\frac{%
G_{0}\left( z-z^{\prime },t\right) }{2\pi \hslash a_{H}\left( 0\right) \sqrt{%
2\mu _{H}\left( t\right) S_{H}^{\left( 0\right) }\left( t\right) }}\exp
\left( \frac{i\mu \sigma }{\hslash s}\int_{0}^{t}H\left( \tau \right) \
d\tau \right)  \label{green3D} \\
&&\quad \times \exp \left( \frac{1}{4i\hslash ^{2}S_{H}^{\left( 0\right)
}\left( t\right) }\left( x-x^{\prime }-\frac{c}{e}\int_{0}^{t}\frac{F\left(
\tau \right) }{H\left( \tau \right) }\ d\tau \right) ^{2}\right)  \notag \\
&&\quad \ \times \exp \left( \frac{\left( S_{H}^{\left( 1\right) }\left(
y,y^{\prime },t\right) \right) ^{2}-4S_{H}^{\left( 0\right) }\left( t\right)
S_{H}^{\left( 2\right) }\left( y,y^{\prime },t\right) }{4iS_{H}^{\left(
0\right) }\left( t\right) }\right)  \notag \\
&&\quad \ \ \times \exp \left( \frac{S_{H}^{\left( 1\right) }\left(
y,y^{\prime },t\right) }{2i\hslash S_{H}^{\left( 0\right) }\left( t\right) }%
\left( x-x^{\prime }-\frac{c}{e}\int_{0}^{t}\frac{F\left( \tau \right) }{%
H\left( \tau \right) }\ d\tau \right) \right)  \notag
\end{eqnarray}%
with%
\begin{equation*}
a_{H}\left( 0\right) =\sqrt{\frac{\hslash }{m\omega _{H}\left( 0\right) }}=%
\sqrt{\frac{\hslash c}{\left\vert e\right\vert H\left( 0\right) }}.
\end{equation*}%
Moreover, the quadratic form in this expression for the $3D$-propagator%
\begin{equation}
Q\left( y,y^{\prime },t\right) =\left( S_{H}^{\left( 1\right) }\left(
y,y^{\prime },t\right) \right) ^{2}-4S_{H}^{\left( 0\right) }\left( t\right)
S_{H}^{\left( 2\right) }\left( y,y^{\prime },t\right)  \label{quadratic1}
\end{equation}
(or the discriminant of (\ref{3D6})) can be rewritten as%
\begin{equation}
Q\left( y,y^{\prime },t\right) =A\left( t\right) y^{2}+B\left( t\right)
yy^{\prime }+C\left( t\right) \left( y^{\prime }\right) ^{2}+D\left(
t\right) y+E\left( t\right) y^{\prime }+L\left( t\right) ,
\label{quadratic2}
\end{equation}%
where%
\begin{eqnarray}
A &=&\frac{1}{2}\frac{\partial ^{2}Q}{\partial y^{2}},\quad \quad \quad B=%
\frac{\partial ^{2}Q}{\partial y\partial y^{\prime }},\quad \quad \quad C=%
\frac{1}{2}\frac{\partial ^{2}Q}{\partial y^{\prime 2}},  \label{quadratic3}
\\
D &=&\left. \frac{\partial Q}{\partial y}\right\vert _{y=y^{\prime
}=0},\quad E=\left. \frac{\partial Q}{\partial y^{\prime }}\right\vert
_{y=y^{\prime }=0},\quad L=\left. Q\right\vert _{y=y^{\prime }=0}.  \notag
\end{eqnarray}%
Our final result is%
\begin{eqnarray}
A\left( t\right) &=&\frac{c^{2}\left( \beta _{H}^{2}\left( t\right) -4\alpha
_{H}\left( t\right) \gamma _{H}\left( t\right) \right) }{e^{2}a_{H}^{2}%
\left( t\right) a_{H}^{2}\left( 0\right) H^{2}\left( 0\right) }+\frac{%
2c\beta _{H}\left( t\right) \delta _{H}^{\left( 1\right) }\left( t\right) }{%
\hslash e\mu _{H}\left( t\right) a_{H}\left( t\right) a_{H}\left( 0\right)
H\left( 0\right) }  \label{A} \\
&&+\frac{\left( \delta _{H}^{\left( 1\right) }\left( t\right) \right) ^{2}}{%
\hslash ^{2}\mu _{H}^{2}\left( t\right) }-\frac{4c\alpha _{H}\left( t\right)
\varepsilon _{H}^{\left( 1\right) }\left( t\right) }{mea_{H}^{2}\left(
t\right) a_{H}^{2}\left( 0\right) H\left( 0\right) }-\frac{2\alpha
_{H}\left( t\right) \kappa _{H}^{\left( 2\right) }\left( t\right) }{\hslash
ma_{H}^{2}\left( t\right) },  \notag
\end{eqnarray}%
\begin{eqnarray}
B\left( t\right) &=&-\frac{2c^{2}\left( \beta _{H}^{2}\left( t\right)
-4\alpha _{H}\left( t\right) \gamma _{H}\left( t\right) \right) }{%
e^{2}a_{H}^{2}\left( t\right) a_{H}^{2}\left( 0\right) H^{2}\left( 0\right) }%
-\frac{2c\beta _{H}\left( t\right) \delta _{H}^{\left( 1\right) }\left(
t\right) }{\hslash e\mu _{H}\left( t\right) a_{H}\left( t\right) a_{H}\left(
0\right) H\left( 0\right) }  \label{B} \\
&&+\frac{4c\alpha _{H}\left( t\right) \varepsilon _{H}^{\left( 1\right)
}\left( t\right) }{mea_{H}^{2}\left( t\right) a_{H}^{2}\left( 0\right)
H\left( t\right) }+\frac{4c\gamma _{H}\left( t\right) \delta _{H}^{\left(
1\right) }\left( t\right) }{\hslash e\mu _{H}\left( t\right) a_{H}^{2}\left(
0\right) H\left( 0\right) }+\frac{2\delta _{H}^{\left( 1\right) }\left(
t\right) \varepsilon _{H}^{\left( 1\right) }\left( t\right) }{\hslash m\mu
_{H}\left( t\right) a_{H}^{2}\left( 0\right) }  \notag \\
&&-\frac{2c\beta _{H}\left( t\right) \varepsilon _{H}^{\left( 1\right)
}\left( t\right) }{mea_{H}\left( t\right) a_{H}^{3}\left( 0\right) H\left(
0\right) }-\frac{2\beta _{H}\left( t\right) \kappa _{H}^{\left( 2\right)
}\left( t\right) }{\hslash ma_{H}\left( t\right) a_{H}\left( 0\right) }, 
\notag
\end{eqnarray}%
\begin{eqnarray}
C\left( t\right) &=&\frac{c^{2}\left( \beta _{H}^{2}\left( t\right) -4\alpha
_{H}\left( t\right) \gamma _{H}\left( t\right) \right) }{e^{2}a_{H}^{2}%
\left( t\right) a_{H}^{2}\left( 0\right) H^{2}\left( t\right) }+\frac{%
2c\beta _{H}\left( t\right) \varepsilon _{H}^{\left( 1\right) }\left(
t\right) }{mea_{H}\left( t\right) a_{H}^{3}\left( 0\right) H\left( t\right) }
\label{C} \\
&&+\frac{\left( \varepsilon _{H}^{\left( 1\right) }\left( t\right) \right)
^{2}}{m^{2}a_{H}^{4}\left( 0\right) }-\frac{4c\gamma _{H}\left( t\right)
\delta _{H}^{\left( 1\right) }\left( t\right) }{\hslash e\mu _{H}\left(
t\right) a_{H}^{2}\left( 0\right) H\left( t\right) }-\frac{4\gamma
_{H}\left( t\right) \kappa _{H}^{\left( 2\right) }\left( t\right) }{2\hslash
ma_{H}^{2}\left( 0\right) },  \notag
\end{eqnarray}%
\begin{eqnarray}
D\left( t\right) &=&\frac{4c^{2}\alpha _{H}\left( t\right) \varepsilon
_{F}^{\left( 0\right) }\left( t\right) }{me^{2}a_{H}^{2}\left( t\right)
a_{H}^{2}\left( 0\right) H\left( t\right) H\left( 0\right) }+\frac{%
2c^{2}\beta _{H}\left( t\right) \varepsilon _{F}^{\left( 0\right) }\left(
t\right) }{me^{2}a_{H}\left( t\right) a_{H}^{3}\left( 0\right) H^{2}\left(
0\right) }  \label{D} \\
&&+\frac{2c\left( \delta _{H}^{\left( 1\right) }\left( t\right) \varepsilon
_{F}^{\left( 0\right) }\left( t\right) -2\delta _{F}^{\left( 0\right)
}\left( t\right) \varepsilon _{H}^{\left( 1\right) }\left( t\right) \right) 
}{\hslash me\mu _{H}\left( t\right) a_{H}^{2}\left( 0\right) H\left(
0\right) }+\frac{2c\alpha _{H}\left( t\right) \kappa _{F}^{\left( 1\right)
}\left( t\right) }{\hslash mea_{H}^{2}\left( t\right) H\left( t\right) } 
\notag \\
&&+\frac{c\beta _{H}\left( t\right) \kappa _{F}^{\left( 1\right) }\left(
t\right) }{\hslash mea_{H}\left( t\right) a_{H}\left( 0\right) H\left(
0\right) }+\frac{\delta _{H}^{\left( 1\right) }\left( t\right) \kappa
_{F}^{\left( 1\right) }\left( t\right) -2\delta _{F}^{\left( 0\right)
}\left( t\right) \kappa _{H}^{\left( 2\right) }\left( t\right) }{\hslash
^{2}m\mu _{H}\left( t\right) }  \notag \\
&&-\frac{2c^{2}\beta _{H}\left( t\right) \delta _{F}^{\left( 0\right)
}\left( t\right) }{\hslash e^{2}\mu _{H}\left( t\right) a_{H}\left( t\right)
a_{H}\left( 0\right) H\left( t\right) H\left( 0\right) }-\frac{4c^{2}\gamma
_{H}\left( t\right) \delta _{F}^{\left( 0\right) }\left( t\right) }{\hslash
e^{2}\mu _{H}\left( t\right) a_{H}^{2}\left( 0\right) H^{2}\left( 0\right) }
\notag \\
&&-\frac{2c\delta _{F}^{\left( 0\right) }\left( t\right) \delta _{H}^{\left(
1\right) }\left( t\right) }{\hslash ^{2}e\mu _{H}^{2}\left( t\right) H\left(
t\right) },  \notag
\end{eqnarray}%
\newline
\begin{eqnarray}
E\left( t\right) &=&\frac{2c^{2}\beta _{H}\left( t\right) \delta
_{F}^{\left( 0\right) }\left( t\right) }{\hslash e^{2}\mu _{H}\left(
t\right) a_{H}\left( t\right) a_{H}\left( 0\right) H^{2}\left( t\right) }+%
\frac{4c^{2}\gamma _{H}\left( t\right) \delta _{F}^{\left( 0\right) }\left(
t\right) }{\hslash e^{2}\mu _{H}\left( t\right) a_{H}^{2}\left( 0\right)
H\left( t\right) H\left( 0\right) }  \label{E} \\
&&+\frac{2c\left( \delta _{F}^{\left( 0\right) }\left( t\right) \varepsilon
_{H}^{\left( 1\right) }\left( t\right) -2\delta _{H}^{\left( 1\right)
}\left( t\right) \varepsilon _{F}^{\left( 0\right) }\left( t\right) \right) 
}{\hslash me\mu _{H}\left( t\right) a_{H}^{2}\left( 0\right) H\left(
t\right) }+\frac{c\beta _{H}\left( t\right) \kappa _{F}^{\left( 1\right)
}\left( t\right) }{\hslash mea_{H}\left( t\right) a_{H}\left( 0\right)
H\left( t\right) }  \notag \\
&&+\frac{2c\gamma _{H}\left( t\right) \kappa _{F}^{\left( 1\right) }\left(
t\right) }{\hslash mea_{H}^{2}\left( 0\right) H\left( 0\right) }+\frac{%
\varepsilon _{H}^{\left( 1\right) }\left( t\right) \kappa _{F}^{\left(
1\right) }\left( t\right) -2\varepsilon _{F}^{\left( 0\right) }\left(
t\right) \kappa _{H}^{\left( 2\right) }\left( t\right) }{\hslash
m^{2}a_{H}^{2}\left( 0\right) }  \notag \\
&&-\frac{4c^{2}\alpha _{H}\left( t\right) \varepsilon _{F}^{\left( 0\right)
}\left( t\right) }{me^{2}a_{H}^{2}\left( t\right) a_{H}^{2}\left( 0\right)
H^{2}\left( t\right) }-\frac{2c^{2}\beta _{H}\left( t\right) \varepsilon
_{F}^{\left( 0\right) }\left( t\right) }{me^{2}a_{H}\left( t\right)
a_{H}^{3}\left( 0\right) H\left( t\right) H\left( 0\right) }  \notag \\
&&-\frac{2c\delta _{F}^{\left( 0\right) }\left( t\right) \delta _{H}^{\left(
1\right) }\left( t\right) }{m^{2}ea_{H}^{4}\left( 0\right) H\left( 0\right) }%
,  \notag
\end{eqnarray}%
\begin{eqnarray}
L\left( t\right) &=&\frac{c^{2}\left( \delta _{F}^{\left( 0\right) }\left(
t\right) \right) ^{2}}{\hslash ^{2}e^{2}\mu _{H}^{2}\left( t\right)
H^{2}\left( t\right) }+\frac{c^{2}\left( \varepsilon _{F}^{\left( 0\right)
}\left( t\right) \right) ^{2}}{m^{2}e^{2}a_{H}^{4}\left( 0\right)
H^{2}\left( 0\right) }+\frac{\left( \kappa _{F}^{\left( 1\right) }\left(
t\right) \right) ^{2}}{4\hslash ^{2}m^{2}}  \label{F} \\
&&+\frac{2c^{2}\delta _{F}^{\left( 0\right) }\left( t\right) \varepsilon
_{F}^{\left( 0\right) }\left( t\right) }{\hslash me^{2}\mu _{H}\left(
t\right) a_{H}^{2}\left( 0\right) H\left( t\right) H\left( 0\right) }+\frac{%
c\delta _{F}^{\left( 0\right) }\left( t\right) \kappa _{F}^{\left( 1\right)
}\left( t\right) }{\hslash ^{2}me\mu _{H}\left( t\right) H\left( t\right) }+%
\frac{c\varepsilon _{F}^{\left( 0\right) }\left( t\right) \kappa
_{F}^{\left( 1\right) }\left( t\right) }{\hslash m^{2}ea_{H}^{2}\left(
0\right) H\left( 0\right) }  \notag \\
&&-\frac{2c^{2}\alpha _{H}\left( t\right) \kappa _{F}^{\left( 0\right)
}\left( t\right) }{\hslash me^{2}a_{H}^{2}\left( t\right) H^{2}\left(
t\right) }-\frac{2c^{2}\beta _{H}\left( t\right) \kappa _{F}^{\left(
0\right) }\left( t\right) }{\hslash me^{2}a_{H}\left( t\right) a_{H}\left(
0\right) H\left( t\right) H\left( 0\right) }-\frac{2c^{2}\gamma _{H}\left(
t\right) \kappa _{F}^{\left( 0\right) }\left( t\right) }{\hslash
me^{2}a_{H}^{2}\left( 0\right) H^{2}\left( 0\right) }  \notag \\
&&-\frac{2c\delta _{H}^{\left( 1\right) }\left( t\right) \kappa _{F}^{\left(
0\right) }\left( t\right) }{\hslash ^{2}me\mu _{H}\left( t\right) H\left(
t\right) }-\frac{2c\varepsilon _{H}^{\left( 1\right) }\left( t\right) \kappa
_{F}^{\left( 0\right) }\left( t\right) }{\hslash m^{2}ea_{H}^{2}\left(
0\right) H\left( 0\right) }-\frac{\kappa _{F}^{\left( 0\right) }\left(
t\right) \kappa _{H}^{\left( 2\right) }\left( t\right) }{\hslash ^{2}m^{2}}.
\notag
\end{eqnarray}%
The details are left to the reader.\smallskip

Finally, by the superposition principle, a general solution of the Cauchy
initial value problem in $\boldsymbol{R}^{3}$ subject to the initial data%
\begin{equation}
\left. \Psi \left( \boldsymbol{r},t\right) \right\vert _{t=0}=\Psi \left( 
\boldsymbol{r},0\right) =\phi \left( x,y,z\right)  \label{sol3Din}
\end{equation}%
has the form%
\begin{equation}
\Psi \left( \boldsymbol{r},t\right) =\int_{\boldsymbol{R}^{3}}G\left( 
\boldsymbol{r},\boldsymbol{r}^{\prime },t\right) \ \Psi \left( \boldsymbol{r}%
^{\prime },0\right) \ dx^{\prime }dy^{\prime }dz^{\prime }.  \label{sol3Dout}
\end{equation}%
This gives us the time evolution operator (\ref{int2}) explicitly for a
motion of a charged particle in a uniform magnetic field and also a
perpendicular electric field with a given projection of the spin $%
s_{z}=\sigma $ in the direction of magnetic field.

\section{Two Examples}

\subsection{Motion in a Constant Magnetic Field}

The simplest case occurs as $H^{\prime }=F=0,$ which implies $\mu
_{H}^{\prime \prime }+\omega _{H}^{2}\mu _{H}=0$ with $\mu _{H}=\sin \left(
\omega _{H}t\right) .$ Thus,%
\begin{eqnarray}
&&\alpha _{H}\left( t\right) =\gamma _{H}\left( t\right) =\frac{1}{2}\cot
\left( \omega _{H}t\right) ,\quad \beta _{H}\left( t\right) =-\frac{1}{\sin
\left( \omega _{H}t\right) },  \label{cmf1} \\
&&\delta _{F}^{\left( 0\right) }=\delta _{H}^{\left( 1\right) }=\varepsilon
_{F}^{\left( 0\right) }=\varepsilon _{H}^{\left( 1\right) }=\kappa
_{F}^{\left( 0\right) }=\kappa _{F}^{\left( 1\right) }=\kappa _{H}^{\left(
2\right) }=0,  \notag
\end{eqnarray}%
and%
\begin{eqnarray}
&&S_{H}^{\left( 0\right) }\left( t\right) =-\frac{1}{m\hslash \omega }\tan
\left( \frac{\omega _{H}t}{2}\right) ,\qquad S_{H}^{\left( 1\right) }\left(
y,y^{\prime },t\right) =-\frac{1}{\hslash }\frac{e}{\left\vert e\right\vert }%
\tan \left( \frac{\omega _{H}t}{2}\right) \left( y+y^{\prime }\right) ,
\label{cmf2} \\
&&S_{H}^{\left( 2\right) }\left( y,y^{\prime },t\right) =\frac{m\omega }{%
2\hslash }\left( \cot \left( \omega _{H}t\right) y^{2}-\frac{2yy^{\prime }}{%
\sin \left( \omega _{H}t\right) }+\cot \left( \omega _{H}t\right) \left(
y^{\prime }\right) ^{2}\right) ,  \notag
\end{eqnarray}%
where the discriminant is%
\begin{equation}
Q\left( y,y^{\prime }\right) =\left( S_{H}^{\left( 1\right) }\left(
y,y^{\prime },t\right) \right) ^{2}-4S_{H}^{\left( 0\right) }\left( t\right)
S_{H}^{\left( 2\right) }\left( y,y^{\prime },t\right) =\frac{1}{\hslash ^{2}}%
\left( y-y^{\prime }\right) ^{2}.  \label{cmf3}
\end{equation}%
Hence, the Green function is given by%
\begin{align}
& G\left( \boldsymbol{r},\boldsymbol{r}^{\prime },t\right) =G_{0}\left(
z-z^{\prime },t\right) \exp \left( \dfrac{i\mu \sigma Ht}{\hslash s}\right)
\ \frac{m\omega _{H}}{4\pi i\hslash \sin \left( \omega _{H}t/2\right) }\ 
\label{cmf4} \\
& \quad \times \exp \left( \dfrac{im\omega _{H}}{4\hslash }\left( \left(
\left( x-x^{\prime }\right) ^{2}+\left( y-y^{\prime }\right) ^{2}\right)
\cot \left( \omega _{H}t/2\right) -2\frac{e}{\left\vert e\right\vert }\left(
x-x^{\prime }\right) \left( y+y^{\prime }\right) \right) \right) .  \notag
\end{align}%
See \cite{Lop:Sus} for more details.

\subsection{A Linear Magnetic Field}

Now consider the case $H\left( t\right) =H_{0}+tH_{1},$ where $H_{0}$ and $%
H_{1}$ are constants. The characteristic equation (\ref{ll12}) becomes a
special case of the Lommel equation \cite{An:As:Ro}, \cite{Ni:Su:Uv}, \cite%
{Sus:Trey}, which can be solved in terms of Bessel functions \cite{Wa} of
orders $\nu =\pm 1/4.$ It then follows that%
\begin{eqnarray}
\mu _{H}\left( t\right) &=&\frac{\pi \left\vert e\right\vert H_{0}^{3/2}}{%
2^{3/2}mcH_{1}}\sqrt{H\left( t\right) }  \label{lmf} \\
&&\times \left( J_{-1/4}\left( \frac{\left\vert e\right\vert H_{0}^{2}}{%
2mcH_{1}}\right) J_{1/4}\left( \frac{\left\vert e\right\vert H^{2}\left(
t\right) }{2mcH_{1}}\right) -J_{1/4}\left( \frac{\left\vert e\right\vert
H_{0}^{2}}{2mcH_{1}}\right) J_{-1/4}\left( \frac{\left\vert e\right\vert
H^{2}\left( t\right) }{2mcH_{1}}\right) \right)  \notag
\end{eqnarray}%
with%
\begin{eqnarray}
\frac{d\mu _{H}\left( t\right) }{dt} &=&\frac{\pi e^{2}}{m^{2}c^{2}H_{1}}%
\left( \frac{H_{0}H\left( t\right) }{2}\right) ^{3/2}  \label{lmfd} \\
&&\times \left( J_{1/4}\left( \frac{\left\vert e\right\vert H_{0}^{2}}{%
2mcH_{1}}\right) J_{3/4}\left( \frac{\left\vert e\right\vert H^{2}\left(
t\right) }{2mcH_{1}}\right) +J_{-1/4}\left( \frac{\left\vert e\right\vert
H_{0}^{2}}{2mcH_{1}}\right) J_{-3/4}\left( \frac{\left\vert e\right\vert
H^{2}\left( t\right) }{2mcH_{1}}\right) \right) .  \notag
\end{eqnarray}%
A general expression of the propagator is given above by (\ref{green3D}). In
the case when $F\equiv 0,$ one can simplify $\delta _{F}^{\left( 0\right)
}=\varepsilon _{F}^{\left( 0\right) }=\kappa _{F}^{\left( 0\right) }=\kappa
_{F}^{\left( 1\right) }\equiv 0.$ We shall elaborate on these and some other
interesting special cases elsewhere.\smallskip

\noindent \textbf{Acknowledgments.\/} The authors are grateful to Professor
Carlos Castillo-Ch\'{a}vez for support and reference \cite{Bet:Cin:Kai:Cas}.
We thank Professors George Andrews, Slim Ibrahim, Hunk Kuiper, Alex Mahalov,
Mizan Rahman, Christian Ringhofer, Svetlana Roudenko, and Hal Smith for
valuable comments. We dedicate this paper to the memory of Professor Basil
Nicolaenko for his significant contributions to the area of nonlinear
partial differential equations, applied mathematics, and related topics.


\begin{thebibliography}{99}
\bibitem{An:As:Ro} G.~E.~Andrews, R.~A.~Askey, and R.~Roy, \textsl{Special
Functions\/}, Cambridge University Press, Cambridge, 1999.

\bibitem{Arrighini:Durante} G.~P.~Arrighini and N.~L.~Durante, \emph{More on
the quantum propagator of a particle in a linear potential\/}, Am. J. Phys. 
\textbf{64} (1996) \#~8, 1036--1041.

\bibitem{Bo:Shi} N.~N.~Bogoliubov and D.~V.~Shirkov, \textsl{Introduction to
the Theory of Quantized Fields\/}, third edition, John Wiley \& Sons, New
York, Chichester, Brisbane, Toronto, 1980.

\bibitem{Beauregard} L.~A.~Beauregard, \emph{Propagators in nonrelativistic
quantum mechanics\/},~Am. J. Phys. \textbf{34} (1966), 324--332.

\bibitem{Bet:Cin:Kai:Cas} L.~M.~A.~Bettencourt, A.~Cintr\'{o}n-Arias,
D.~I.~Kaiser, and C.~Castillo-Ch\'{a}vez, \emph{The power of a good idea:
Quantitative modeling of the spread of ideas from epidemiological models\/}%
,~Phisica~A \textbf{364} (2006), 513--536.

\bibitem{Brown:Zhang} L.~S.~Brown and Y.~Zhang, \emph{Path integral for the
motion of a particle in a linear potential\/}, Am. J. Phys. \textbf{62}
(1994) \# 9, 806--808.

\bibitem{FeynmanPhD} R.~P.~Feynman, \emph{The Principle of Least Action in
Quantum Mechanics\/}, Ph.~D. thesis, Princeton University, 1942; reprinted
in: \textsl{\textquotedblleft Feynman's Thesis -- A New Approach to Quantum
Theory\textquotedblright\/}, (L.~M.~Brown, Editor), World Scientific
Publishers, Singapore, 2005, pp.~1--69.

\bibitem{Feynman} R.~P.~Feynman, \emph{Space-time approach to
non-relativistic quantum mechanics\/}, Rev. Mod. Phys. \textbf{20}
(1948)~\#~2, 367--387; reprinted in: \textsl{\textquotedblleft Feynman's
Thesis -- A New Approach to Quantum Theory\textquotedblright\/},
(L.~M.~Brown, Editor), World Scientific Publishers, Singapore, 2005,
pp.~71--112.

\bibitem{Feynman49a} R.~P.~Feynman, \emph{The theory of positrons\/}, Phys.
Rev. \textbf{76} (1949)~\#~6, 749--759.

\bibitem{Feynman49b} R.~P.~Feynman, \emph{Space-time approach to quantum
electrodynamics\/}, Phys. Rev. \textbf{76} (1949)~\#~6, 769--789.

\bibitem{Fey:Hib} R.~P.~Feynman and A.~R.~Hibbs, \textsl{Quantum Mechanics
and Path Integrals\/}, McGraw--Hill, New York, 1965.

\bibitem{Flu} S.~Fl\"{u}gge, \textsl{Practical Quantum Mechanics\/},
Springer--Verlag, Berlin, 1999.

\bibitem{Gottf:T-MY} K.~Gottfried and T.-M.~Yan, \textsl{Quantum Mechanics:
Fundamentals\/}, second edition, Springer--Verlag, Berlin, New York, 2003.

\bibitem{Haah:Stein} D.~R.~Haaheim and F.~M.~Stein, \emph{Methods of
solution of the Riccati differential equation\/}, Mathematics Magazine 
\textbf{42} (1969)~\#2, 233--240.

\bibitem{Holstein97} B.\ R.\ Holstein, \emph{The linear potential
propagator\/}, Am. J. Phys. \textbf{65} (1997)~\#5, 414--418.

\bibitem{Holstein} B.\ R.\ Holstein, \emph{The harmonic oscillator
propagator\/}, Am. J. Phys. \textbf{67} (1998)~\#7, 583--589.

\bibitem{Howland} J.~Howland, \emph{Scattering theory for Hamiltonians
periodic in time\/}, Indiana Univ. Math. J. \textbf{28} (1979)~\# 3,
471--494.

\bibitem{Jack} J.~D.~Jackson, \textsl{Classical Electrodynamics}, Second
Edition, John Wiley \& Sons, New York, 1975.

\bibitem{Jafaev} D.~R.~Jafaev, \emph{Wave operators for the Schr\"{o}dinger
equation\/}, [in Russian] Teoret. Mat. Fiz. \textbf{45} (1980)~\#2, 224--234.

\bibitem{La:Lif} L.~D.~Landau and E.~M.~Lifshitz, \textsl{Quantum Mechanics:
Nonrelativistic Theory\/}, Pergamon Press, Oxford, 1977.

\bibitem{Lop:Sus} R.~M.~Lopez and S.~K.~Suslov, \emph{The Cauchy problem for
a forced harmonic oscillator\/}, arXiv:0707.1902v8 [math-ph] 27 Dec 2007.

\bibitem{Maslov:Fedoriuk} V.~P.~Maslov and M.~V.~Fedoriuk, \textsl{%
Semiclassical Approximation in Quantum Mechanics\/}, Reidel, Dordrecht,
Boston, 1981.

\bibitem{Me:Co:Su} M.~Meiler, R.~Cordero--Soto, and S.~K.~Suslov, \emph{%
Solution of the Cauchy problem for a time-dependent Schr\"{o}dinger
equation\/}, arXiv: 0711.0559v4 [math-ph] 5 Dec 2007.

\bibitem{Merz} E.~Merzbacher, \textsl{Quantum Mechanics\/}, third edition,
John Wiley \& Sons, New York, 1998.

\bibitem{Mes} A.~Messia, \textsl{Quantum Mechanics\/}, two volumes, Dover
Publications, New York, 1999.

\bibitem{Naibo:Stef} V.~Naibo and A.~Stefanov, \emph{On some Schr\"{o}dinger
and wave equations with time dependent potentials\/}, Math. Ann. \textbf{334}
(2006) \# 2, 325--338.

\bibitem{Nardone} P.~Nardone, \emph{Heisenberg picture in quantum mechanics
and linear evolutionary systems\/}, Am. J. Phys. \textbf{61} (1993) \# 3,
232--237.

\bibitem{Ni:Su:Uv} A.~F.~Nikiforov, S.~K.~Suslov, and V.~B.~Uvarov, \textsl{%
Classical Orthogonal Polynomials of a Discrete Variable\/},
Springer--Verlag, Berlin, New York, 1991.

\bibitem{Ni:Uv} A.~F.~Nikiforov and V.~B.~Uvarov, \textsl{Special Functions
of Mathematical Physics\/}, Birkh\"{a}user, Basel, Boston, 1988.

\bibitem{Palio:Mead} J.~D.~Paliouras and D.~S.~Meadows, \textsl{Complex
Variables for Scientists and Engineers\/}, second edition, Macmillan
Publishing Company, New York and London, 1990.

\bibitem{Rain} E.~D.~Rainville, \textsl{Special Functions\/}, The Macmillan
Company, New York, 1960.

\bibitem{Rainville} E.~D.~Rainville, \textsl{Intermediate Differential
Equations\/}, Wiley, New York, 1964.

\bibitem{Robinett} R.~W.~Robinett, \emph{Quantum mechanical time-development
operator for the uniformly accelerated particle\/}, Am. J. Phys. \textbf{64}
(1996) \#6, 803--808.

\bibitem{Rod:Schlag} I.~Rodnianski and W.~Schlag, \emph{Time decay for
solutions of Schr\"{o}dinger equations with rough and time-dependent
potentials}\textit{\/}, Invent. Math. \textbf{155} (2004)~\# 3, 451--513.

\bibitem{Schlag} W.~Schlag, \emph{Dispersive estimates for Schr\"{o}dinger
operators: a survay}\textit{\/}, arXiv: math/0501037v3 [math.AP] 10 Jan 2005.

\bibitem{SteinHarm} E.~M.~Stein, \textsl{Harmonic Analysis: Real-Variable
Methods, Orthogonality, and Oscillatory Integrals\/}, Princeton University
Press, Princeton, New Jersey, 1993.

\bibitem{Styer} D.~Styer, \textsl{Additions and Corrections to Feynman and
Hibbs\/}, see an updated version on the author's web site:\linebreak
http://www.oberlin.edu/physics/dstyer/FeynmanHibbs/errors.pdf.

\bibitem{Sus:Trey} S.~K.~Suslov and B.~Trey, \emph{The Hahn polynomials in
the nonrelativistic and relativistic Coulomb problems\/}, J. Math. Phys. 
\textbf{49} (2008) \#1, published on line 22 January 2008, URL:
http://link.aip.org/link/?JMP/49/012104.

\bibitem{Thomber:Taylor} N.~S.~Thomber and E.~F.~Taylor, \emph{Propagator
for the simple harmonic oscillator\/},~Am. J. Phys. \textbf{66} (1998) \#
11, 1022--1024.

\bibitem{Wa} G.~N.~Watson, \textsl{A Treatise on the Theory of Bessel
Functions\/}, Second Edition, Cambridge University Press, Cambridge, 1944.

\bibitem{Yajima} K.~Yajima, \emph{Scattering theory for Schr\"{o}dinger
equations with potentials periodic in time}\textit{\/}, J.~Math. Soc. Japan 
\textbf{29} (1977) \# 4, 729--743.
\end{thebibliography}
\end{document}